\documentstyle[e-e-ijmpa,twoside]{article}
\input psfig.sty

\begin{document}

\normalsize\textlineskip
\pagestyle{empty}

\title{LARGE EXTRA DIMENSIONS AT LINEAR COLLIDERS}

\author{K. SRIDHAR 
\footnote{sridhar@theory.tifr.res.in}
}

\address{Department of Theoretical Physics,\\
Tata Institute of Fundamental Research,\\
Homi Bhabha Road, Mumbai 400 005, INDIA}

\maketitle\abstracts{In this talk, I first present the motivation for
theories wherein the extra spacetime dimensions can be compactified to 
have large magnitudes. In particular, I discuss the Arkani-Hamed, Dimopoulos,
Dvali (ADD) scenario. I present the constraints that have been derived
on these models from current experiments and the expectations from future
colliders. I concentrate particularly on the possibilities of probing
these extra dimensions at future linear colliders.  }

\setcounter{footnote}{0}

\vspace*{1pt}\textlineskip        
\section{Introduction to the Kaluza-Klein theory}
\noindent Very soon after the formulation of General Relativity, and its
success in giving gravity a geometrical meaning, attempts at 
unifying electromagnetism with gravity were made. These theories,
known as Kaluza-Klein theories, attempted to obtain gravity and
electromagnetism from the geometry of an underlying higher-dimensional
theory. 
Non-observation of these extra dimensions implies that these have to 
be compactified to sizes which are unobservably small.
As a simple example, consider a scalar field in 5 dimensions,
$\phi (x, y)$ where $x$ is the 4-dimensional space-time co-ordinate
and $y$ is the 5th dimension. Assume that the fifth dimension is
compactified to a circle with radius $R$, where $R$ is independent
of $x$. 
The 5-dimensional field can be expanded in a Fourier series as
\begin{equation} 
\phi (x, y) = \sum_{n=-\infty}^{\infty} \phi_n (x) {\rm exp} (iny/R) ,
\end{equation} 
where $n$ is an integer and $\phi_n (x)$ are 4-dimensional fields.
Substituting the above expansion in the 5-dimensional Klein-Gordon equation
which $\phi (x, y)$ satisfies, one can show that one ends up with an
infinite number of equations in 4 dimensions, one for each $\phi_n (x)$
and with a mass $ \vert n \vert / R$ for every mode n. These modes
are called $pyrgons$.
The definition of spin in $D$ dimensions depends on the $D$-dimensional
Lorentz symmetry. The light-cone symmetry that leaves the motion of a
massless particle unchanged in $D$ dimensions is $SO (D-2)$ and the
$D$-dimensional helicity corresponds to the representations of $SO (D-2)$.
A given Kaluza-Klein level in $D=4+m$ dimensions has one spin-2 state,
$(m-1)$ spin-1 states and $m(m-1)/2$ spin-0 states. A higher dimensional
field, therefore, unifies different fields of different masses and
spins in 4 dimensions. 

The problem, however, is these extra dimensions
are not observed. In fact, if the periodicity in the fifth dimension
is related to the quantisation of the electric charge then the length
of the extra dimension turns out to be of the order of $10^{-30}$ cm
which is only somewhat bigger than the Planck length, making the 
hypothesis of extra dimensions untestable in any experiment. This
is an unattractive feature of these theories, nevertheless the idea
that all interactions are the consequence of space-time symmetries
is so attractive that there have been vigorous attempts to generalise
the attempt of Kaluza and Klein to include other interactions using
more complicated compactification schemes.

\vspace*{1pt}\textlineskip        
\section{Large Extra Dimensions}
Recently, new incarantaions of Kaluza-Klein theories have been
discussed in the literature which can be a way of getting around
the so-called hierarchy problem. What is the hierarchy problem?
The Standard Model (SM) has proved enormously successful in providing a 
description of particle physics upto energy scales probed by current
experiments, which is in the region of several hundred GeV. In the SM,
however, one assumes that effects of gravity can be neglected, because
the scale where the effects of gravity become large i.e. the Planck
scale ($M_P = 1.2 \times 10^{19}$~GeV) is vastly different from the TeV scale.
The separation between the TeV scale and the Planck scale is what 
manifests itself as the hierarchy problem, whose solution has become
one of the foci of the search for the correct physics beyond the SM.
This problem is exacerbated in traditional unification scenarios:
the scale of grand-unification is of the order of $10^{16}$~GeV and
again implies a huge desert. Further, in spite of the unification scale
being so close to the Planck scale, traditional unification models make
no reference whatsoever to gravity. 

Recent advances in the understanding of the strong-coupling regime of string 
theories has led to a major paradigm shift. The tool that has made it 
possible to understand the strong-coupling regime is duality. This duality,
which is quite similar to the concept of duality in field theories, relates
a theory at weak coupling to another theory at strong coupling. In field
theories, this relationship also entails an electric/magnetic duality
where duality takes a theory of weakly coupled point-like electric
charges (and strongly coupled magnetic charges) to one with magnetic
charges that are weakly coupled and pointlike. The strongly coupled theory
maps on to the weakly coupled theory in which the basic quanta carry
magnetic charges. In field theory, therefore, the duality multiplets
include the elementary quanta which are pointlike and solitonic modes
which are extended configurations. The situtation in string theory is
more complicated where, in addition to the elementary strings, the
spectrum of particles includes solitonic objects which are called D-branes.
These are best thought of as topological defects of varying dimensionality:
a D-brane is a dynamical $D+1$-dimensional surface.
The weak coupling string theory is not sensitive to these modes because
they are very heavy compared to the stringy modes but, on the other hand, in
the dual theory they become lighter and so in the dual theory it is
best to think of the D-branes as the elementary quanta. An interesting
feature of the $D$-branes is that they act as surfaces on which open
strings end. 

Now let us consider a theory with 3-branes or $3+1$-dimensional 
hypersurfaces which are embedded in a $D$-dimensional spacetime.
This theory would have typically open and closed strings along with
the 3-branes. The gauge particles, which correspond to the open strings,
will end on the 3-branes while the gravitons, which correspond to
the closed strings, are not restricted to lie on the 3-brane. This
implies that the gauge particles (i.e. the SM particles) are confined
to the 3-brane or the 3+1 dimensional surface and only the gravitons
are free to propagate in the full $D$ dimensions. As usual, the extra $D-4$
dimensions have to be compactified to obtain the $3+1$ dimensional
theory \cite{string}. But, since these extra dimensions are 
only `seen' by gravity,
these need not be compactified to length scales which are of the
order of $M_{P}^{-1}$ but it is can be arranged that $n$ of these 
extra dimensions
are compactified to a common scale $R$ which is relatively large,
while the remaining dimensions are compactified to much smaller
length scales which are of the order of the inverse Planck scale.
In this context, the idea of large extra dimensions was first
discussed by Arkani-Hamed, Dimopoulos and Dvali \cite{dimo}
and is referred to as the ADD scenario, though earlier attempts
at making the extra dimensions large have been made \cite{anto}.
The relation between 
the scales in $4+n$ dimensions and in $4$ dimensions is given by \cite{dimo}
\begin{equation} 
M^2_{\rm P}=M_{S}^{n+2} R^n ~,
\label{e1} 
\end{equation} 
where $M_S$ is the low-energy effective string scale. This equation has
the interesting consequence that we can choose $M_S$ to be of the order
of a TeV and thus get around the hierarchy problem
\footnote{A more recent scenario due to Randall and Sundrum \cite{rs}
is, in fact, a better way of handling the hierarchy problem. The
phenomenlogy of this scenario, however, is not very different than the ADD
scenario discussed in this paper.}.
For such a value of
$M_S$, it follows that $R=10^{32/n -19}$~m, and so we find that $M_S$
can be arranged to be a TeV for any value $n > 1$. Effects of non-Newtonian
gravity can become apparent at these surprisingly low values of energy.
For example, for $n=2$ the compactified dimensions
are of the order of 1 mm, just below the experimentally tested region
for the validity of Newton's law of gravitation and within the possible
reach of ongoing experiments \cite{gravexp}. 

\vspace*{1pt}\textlineskip        
\section{The View from the Braneless End: The Low-Energy Effective Theory}
Below the scale $M_S$ the following effective picture emerges \cite{sundrum}, 
\cite{grw}, \cite{hlz}: 
there are the Kaluza-Klein states, in addition to the usual
SM particles. The graviton corresponds to a tower of Kaluza-Klein states
which contain spin-2, spin-1 and spin-0 excitations. The spin-1
modes do not couple to the energy-momentum tensor and their
couplings to the SM particles in the low-energy effective
theory are not important. The scalar modes couple to the trace
of the energy-momentum tensor, so they do not couple to massless
particles. Other particles related to brane dynamics 
(for example, the $Y$ modes which are related to the
deformation of the brane) have effects which are subleading, compared to
those of the graviton. The only states, then, that contribute 
are the spin-2 Kaluza-Klein states. These
correspond to a massless graviton in the $4+n$ dimensional theory,
but manifest as an infinite tower of massive gravitons in the low-energy
effective theory. For graviton momenta smaller than the scale $M_S$, the
effective description reduces to one where the gravitons in the bulk 
propagate in the flat background and couple to the SM fields which live
on the brane via a (four-dimensional) induced metric $g_{\mu \nu}$. 
Starting from a linearized gravity Lagrangian
in $n$ dimensions, the four-dimensional interactions can be derived after
a Kaluza-Klein reduction has been performed. The interaction of the SM 
particles with the graviton, $G_{\mu\nu}$, can be derived from 
the following Lagrangian:
\begin{equation} 
{\cal L}=-{1 \over \bar M_P} G_{\mu \nu}^{(j)}T^{\mu\nu} ~,
\label{e2} 
\end{equation} 
where $j$ labels the Kaluza-Klein mode and $\bar M_P=M_P/\sqrt{8\pi}$,
and $T^{\mu\nu}$ is the energy-momentum tensor.

In view of the
fact that the effective Lagrangian given in Eq.~\ref{e2} is suppressed
by $1/\bar M_P$, it may seem that the effects at colliders will be hopelessly
suppressed. However, in the case of real graviton production, the phase
space for the Kaluza-Klein modes cancels the dependence on $\bar M_P$ 
and, instead, provides a suppression of the order of $M_S$. For the
case of virtual production, we have to sum over the whole tower of 
Kaluza-Klein states and this sum when properly evaluated \cite{hlz}, \cite{grw}
provides the correct order of suppression ($\sim M_S$). The summation
of time-like propagators and space-like propagators yield exactly the
same form for the leading terms in the expansion of the sum \cite{hlz}
and this shows that the low-energy effective theories for the $s$ and 
$t$-channels are equivalent.

\vspace*{1pt}\textlineskip        
\section{The Experimental Constraints }
There have been several studies exploring the consequences of the
above effective Lagrangian for particle phenomenology and astrophysics.
Production of gravitons giving rise to characteristic missing energy or 
missing $p_T$ signatures at $e^+ e^-$ or hadron colliders have been
studied resulting in bounds on $M_S$ which are around 500 GeV to 
1.2 TeV at LEP2 \cite{mpp}, \cite{keung} 
and around 600 GeV to 750 GeV at Tevatron 
\cite{mpp}. Production of gravitons at the Large Hadron Collider (LHC) and 
in high-energy $e^+ e^-$ collisions at the Next Linear Collider (NLC) have 
also been considered. Virtual effects of graviton exchange in dilepton 
production at Tevatron yields a bound of around 950 GeV \cite{hewett} to 
1100 GeV \cite{gmr} on $M_S$, in $t \bar t$ production at Tevatron a bound of 
about 650 GeV is obtained while at the LHC this process can be used to explore 
a range of $M_S$ values upto 4~TeV \cite{us}. Virtual effects in 
deep-inelastic scattering at HERA put a bound of 550 GeV on $M_S$\cite{us2}, 
while from jet production at the Tevatron strong bounds of about 1.2 TeV are 
obtained \cite{us3}. Pair production of gauge bosons and fermions in $e^+ e^-$ 
collisions at LEP2 \cite{rizzo}, \cite{agashe}, \cite{soni} can 
probe values of $M_S$ upto 
0.6 TeV. Other processes studied include associated production of gravitons 
with gauge bosons and virtual effects in gauge boson pair production at hadron 
colliders \cite{balazs}, \cite{cheung}. Higgs production \cite{rizzo2}, 
\cite{xhe} 
and electroweak precision observables \cite{precision} in the light of this 
new physics have also been discussed. 
Astrophysical constraints, like bounds from energy loss for supernovae cores, 
have also been discussed \cite{astro}. In general, the processes which involve
real production of gravitons give stronger constraints for $n=2$ than the
processes involving virtual exchange of gravitons but the advantage of the
virtual processes is that the bounds obtained from them have a mild $n$
dependence whereas the bounds from real production processes fall rapidly
with increasing $n$.

\vspace*{1pt}\textlineskip        
\section{The Linear Collider and Extra Dimensions}
The Next Linear Collider (NLC) is an ideal testing ground of the SM and 
a very effective probe of possible physics that may lie beyond the SM. 
The collider is planned to be operated in the $e^+ e^-$, $e^- e^-$, 
$\gamma \gamma$ and the $e \gamma$ modes. For operation in the latter
two modes, the photons are produced in the Compton 
back-scattering of a highly monochromatic low-energy laser beam off a 
high energy electron beam \cite{nlc}. Control over the $e^-$ and laser 
beam parameters allow for control over the parameters of the $\gamma \gamma$ 
and $e \gamma$ collisions. The physics potential of the NLC is manifold and 
the collider is expected to span several steps of $e e$ energy between 
500 GeV and 1.5 TeV. The experiments at the NLC also provide a great degree 
of precision because of the relatively clean initial state, and indeed the 
degree of precision can be enhanced by using polarised initial beams. 

We first discuss the case of $e^+ e^-$ collisions. The
production of gravitions in $e^+ e^-$ collisions at the NLC for a $\sqrt{s}$
of 1 TeV and 100 fb${}^{-1}$ luminosity has been studied \cite{mpp}. 
Bounds on $M_S$ which are around 3-7 TeV are obtained (for $n$ between 
2 and 6). 
Virtual effects of graviton exchange in fermion pair production at the
NLC \cite{hewett} can also give strong bounds of upto 5~TeV for a $\sqrt{s}=
1$~TeV. These bounds can be enhanced by studying the angular distributions
instead of looking at the integrated cross-sections.
Virtual effects of graviton exchange in gauge boson pair production at the
NLC have also been studied \cite{lee}, \cite{rizzo} and lead to similarly strong
bounds.

In the $e^-e^-$ mode, M\" oller scattering $e^-e^- \rightarrow  e^-e^- $ 
may be used to study the virtual exchange of gravitons and this process
is similar to the process $e^+e^- \rightarrow e^+ e^-$ as far as the
gaviton exchange contribution is concerned. The $e^- e^-$ mode, in fact,
is advantageous in that it provides a initial state even cleaner than
in $e^+ e^-$. The $e^- e^-$ initial state can also be polarised to a 
greater degree. However, the major advantage that the the $e^+ e^-$
mode has over the $e^- e^-$ mode is that there are several $f \bar f$ states
that are accessible and by summing over all these states the bound
can be significantly improved. 

If operated in a mode where there laser-back scattering used,
both the $e^+ e^-$ and the $e- e^-$ colliders can be used to
study $\gamma \gamma$ and $e \gamma$ scattering processes.
As an example consider the effects of large extra dimensions in top 
production in photon-photon collisions at the NLC, spanning the energy 
range between 500 GeV and 1.5 TeV \cite{us4}. 

The basic scattering is described by a $\gamma \gamma$ scattering
subprocess, with each $\gamma$ resulting from the electron-laser back
scattering. The energy of the back-scattered photon, $E_\gamma$, follows
a distribution characteristic of the Compton scattering process and
can be written in terms of the dimensionless ratio $x=E_\gamma/E_e$.
The subprocess cross-section is convoluted with the luminosity functions, 
$f^i_\gamma (x)$, which provide information on the photon flux produced in 
Compton scattering of the electron and laser beams. 
The cross-section for the $\gamma \gamma \rightarrow t \bar t$ 
process has the usual $t$- and $u$-channel SM contributions, but in
addition, we also have the $s$-channel exchange of virtual spin-2
Kaluza-Klein particles. The $2\sigma$ limits that we obtain for 
$\sqrt{s}= 500,\ 1000,\ 1500$~GeV are 1600, 4000 and 5400 GeV, respectively.
Tighter cuts on the rapidity can be used to improve the bounds
significantly. The use of polarisation enhances the bounds on
$M_S$ quite significantly by several 100 GeV in each case. The
other processes that have been studied in the case of $\gamma \gamma$
collisions are gauge-boson pair production \cite{davoudiasl}, 
\cite{rizzo} and dijet 
production \cite{us5}.

Finally, the effects of extra dimensions in $e\gamma$ collisions
has also been studied. The virtual exchange of gravitons in the $e \gamma
\rightarrow e \gamma$ Compton scattering yields bounds in the range
of 5 TeV \cite{davoudiasl2} and the real production of gravitons $via$ 
$e \gamma \rightarrow e G$ also yields bounds in the region of around 5 TeV
\cite{us6}.

\vspace*{1pt}\textlineskip        
\section{Conclusions}
\noindent The possibility that extra dimensions could be compactified
to sizes as large as a millimeter, and consequently have effects of
quantum gravity in the TeV range, has led to several exciting investigations
of these effects at high energy colliders. In this talk, I have essentially
summarised the motivation for expecting these extra dimensions to be
large, discussed the low-energy effective theory and then reviewed the
bounds on the string scale $M_S$ that have been obtained from present
experiments and the values of $M_S$ that will be probed in future
colliders. In particular, I have discussed these studies made for a future
linear collider and conclude that these theories can be probed to very large 
values of effective string scale $M_S$.

\vspace*{1pt}\textlineskip

\end{document}